\documentclass[american,nofootinbib]{revtex4-2}

\usepackage[T1]{fontenc}
\usepackage[utf8]{inputenc}
\usepackage{xcolor}
\usepackage{babel}
\usepackage{amsmath}
\usepackage{graphicx}
\usepackage{soul}
\usepackage{makecell}
\usepackage{bm}

\usepackage[pdfusetitle,
 bookmarks=true,bookmarksnumbered=false,bookmarksopen=false,
 breaklinks=false,pdfborder={0 0 1},backref=false,colorlinks=true]
 {hyperref}
\hypersetup{
 urlcolor=blue, citecolor=blue, linkcolor=blue}

\newcommand{\bmbhat}{\hat{\bm b}}
\newcommand{\bmnhat}{\hat{\bm n}}
\newcommand{\bmxhat}{\hat{\bm x}}
\newcommand{\bmyhat}{\hat{\bm y}}
\newcommand{\bmzhat}{\hat{\bm z}}

\newcommand{\bmu}{\bm u}
\newcommand{\bmOmega}{\bm \Omega}
\newcommand{\bmB}{\bm B}
\newcommand{\bmE}{\bm E}
\newcommand{\bmJ}{\bm J}
\newcommand{\bmnabla}{\bm \nabla}

\newcommand{\drm}{\mathrm d}

\providecommand{\tabularnewline}{\\}

\makeatother
\begin{document}

%%%% Article title to be placed here
\title{Self-organisation through layering of $\beta$-plane like turbulence in
plasmas and geophysical fluids}

\author{P. L. Guillon$^{1,2}$, G. Dif-Pradalier$^{3}$, Y. Sarazin$^{3}$,
D. W. Hughes$^{4}$, Ö. D. Gürcan$^{1}$}

%%%%%%%%% Insert author address here
\affiliation{$^{1}$Laboratoire de Physique des Plasmas, CNRS, Ecole Polytechnique,
Sorbonne Université, Université Paris-Saclay, Observatoire de Paris,
F-91120 Palaiseau, France}
\affiliation{$^{2}$ENPC, Institut Polytechnique de Paris, 77455 Champs-sur-Marne, Marne-la-Vallée cedex 2, France}
\affiliation{$^{3}$CEA, IRFM, F-13108 Saint-Paul-lez-Durance, France}
\affiliation{$^{4}$School of Mathematics, University of Leeds, Leeds LS2 9JT, UK}

%%%% Abstract text to be placed here %%%%%%%%%%%%
\begin{abstract}
Staircase formation and layering is studied in simplified, potential vorticity conserving models
of plasmas and geophysical fluids, by investigating turbulent self-organisation and nonlinear
saturation with different mechanisms of free energy production --- forcing or linear instability --- and with standard or modified zonal flow responses. To this end, staircase formation in both the standard and modified Charney-Hasegawa-Mima equations with stochastic forcing, along with two different simple instability driven models --- one from a plasma and from a geophysical context --- are studied
and compared.
In these studies, it is observed that $\beta$-plane turbulence that does not distinguish between zonal and non-zonal perturbations (i.e., standard zonal response) gradually forms large-scale, elliptic zonal structures that merge progressively, regardless of whether it is driven by forcing (though it should be weak and slow enough to allow wave couplings) or by the baroclinic instability, using for example a two-layer model.
Conversely, the plasma system, with its modified zonal response, can rapidly form straight, stationary jets of well-defined size, again regardless of the way it is driven: by stochastic forcing or by the dissipative drift-wave instability.  Furthermore, the instability-driven plasma system exhibits a subcritical zonal bifurcation, which resembles a phase transition, between a zonal flow dominated state and an eddy dominated state. In both states, saturation is possible without large-scale friction.
\end{abstract}
%%%%%%%%%%%%%%%%%%%%%%%%%%%

\maketitle
\section{Introduction}\label{sec:Introduction}
The similarities between basic descriptions of turbulence in geophysical fluids and tokamak
plasmas have been known for a while \cite{horton:94}.
They are both quasi-two-dimensional systems, which become so by the
effects of gravity and strong planetary rotation in one case, and
the confining background magnetic field in the other. Two key elements
of this similarity are the fact that the Coriolis force and the Lorentz
force have the same structures, i.e., they are the cross product of a velocity and a frequency (Earth's rotation or cyclotron frequency in the electrostatic limit) along a normal direction defined by hydrostatic balance or MHD equilibrium. Both systems are in static equilibrium, on top of which fluctuations may appear. They also exhibit 
significant large-scale inhomogeneities, thereby resulting in very 
anisotropic turbulence, characterised by wave-like dynamics, whose nonlinear 
modulations then favour the formation of zonal flows (ZFs) \cite{diamond:05,gurcan:15} 
and layering of the initial inhomogeneities \cite{Rhines:1982,dif-pradalier:15}.

Even though the physical details of the two problems are vastly different, with some of these differences being of great importance for the goals of those specific disciplines, in their simplest form, the turbulence
in both systems can be described using a common framework, represented
by the so-called Charney-Hasegawa-Mima (CHM) equations \cite{charney:48,hasegawa:1978}.
In this framework, one can define a conserved quantity, known as the
potential vorticity (PV), whose gradients are responsible for the
corresponding waves --- Rossby waves for geophysical fluids and drift
waves for plasmas --- and whose non-linear evolution gives rise to
$\beta$-plane turbulence \cite{rhines:75}.

\begin{figure}[b]
\centering{}\includegraphics[width=\textwidth]{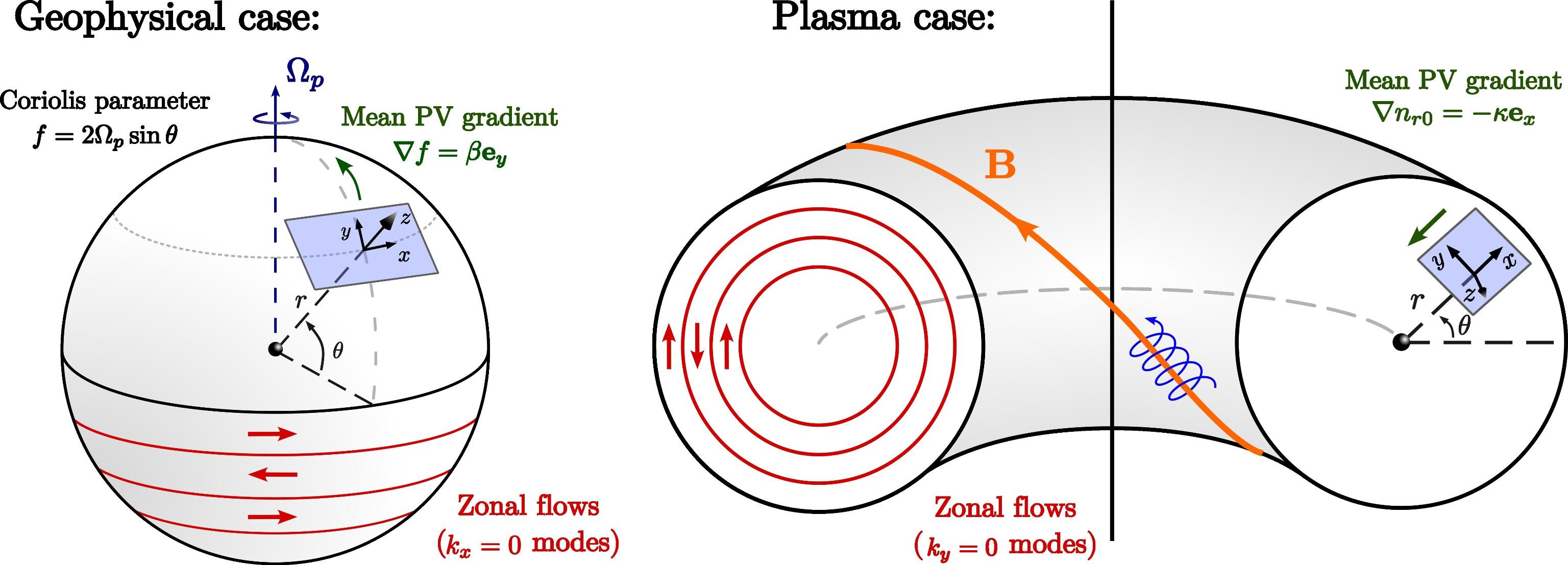}\caption{Summary of the various geometries and conventions used in the geophysical
and plasma problems. The angle $\theta$ is the latitude in the geophysical case, and the poloidal direction in the plasma case. Note that zonal flows correspond to $k_x=0$ modes in the geophysical case, and to $k_y=0$ in the plasma case. In tokamaks, the actual magnetic field lines are wrapped along the torus (orange), but in the minimal models we consider a uniform, constant magnetic field along $\bmzhat$.}
\label{fig:Geometries} 
\end{figure}

In this simple form, the PV can be `inverted' \cite{mcintyre:00,mcintyre:08},
or written as a function of the velocity field, which can in turn
be written in terms of the stream function (or the electrostatic potential
for the plasma case, which has the same role apart from its units),
thus closing the loop. However, the simplest $\beta$-plane models
do not generate turbulence spontaneously and have to be driven
by external forcing. This is usually justified in the geophysical
fluid dynamics (GFD) context by arguing that perturbations due to different physical mechanism, such as the baroclinic instability,
inject energy at small scales, which then piles up towards larger
scales for which the $\beta$-plane approximation is relevant.
Unfortunately in the plasma case, there does not seem to be such scale
separation between where, for example, a drift instability would inject
its energy, and the scales at which the drift wave turbulence would
appear, except in rather unlikely scenarios, such as the energy from
an electron instability (e.g., the electron temperature gradient driven
mode \cite{horton:1988}) somehow cascading towards large scales
to generate drift wave turbulence. Note that for the geophysical problem, the characteristic length scale 
of the perturbations tends to be smaller than the characteristic scale of the underlying dynamics --- the Rossby deformation radius --- whereas for the plasma problem, it tends to be larger than the corresponding scale --- the Larmor radius. In table~\ref{table:sum1}, we give a summary of the analogy between the geophysical and plasma problems. The coordinate system of both problems is illustrated in figure~\ref{fig:Geometries}. In particular, note that the $x$ and $y$ directions are somewhat interchanged in the two problems, $x$ ($y$) being the periodic direction in the geophysical (plasma) context. Also, as discussed later, instabilities arise owing to gradients in the $y$ ($x$) direction in the GFD (plasma) context.

\begin{table}[htbp]
\begin{tabular*}{1\textwidth}{@{\extracolsep{\fill}}lcc}
\hline 
\textbf{Physical context} & \textbf{Geophysical problem} & \textbf{Plasma problem}\tabularnewline
\hline 
Driving force/vector field & Coriolis/planetary rotation $\bmOmega_p$ & Lorentz/magnetic field $\bmB$ \tabularnewline
PV gradient direction & $y$ (south-north) & $x$ (radial) \tabularnewline
Zonal direction & $x$ (west-east) & $y$ (poloidal)\tabularnewline
PV gradient & Planetary vorticity $\beta=\drm f/\drm y$ & Plasma density $\kappa=-\drm n/\drm x$\tabularnewline
Characteristic length & \makecell{Rossby deformation radius \\ $R=\sqrt{gh_0}/f_{0}$} & \makecell{Ion sound Larmor radius \\ $\rho_{s}=c_{s}/\Omega_{i}$} \tabularnewline
Small parameter & Rossby number $U/(Lf_0)$ & $\rho_s/L_n$
\tabularnewline
Advective velocity & Geostrophic flow Eqn. \eqref{eq:geob} & $\bmE \times \bmB$ drift Eqn. \eqref{eq:ExB} \tabularnewline
Scalar Potential & Stream-function (Layer height) $\psi$ & Electrostatic potential $\Phi$ \tabularnewline
\hline
\textbf{$\beta$-plane model} & \textbf{Charney-Hasegawa-Mima} & \textbf{Generalised Hasegawa-Mima} \tabularnewline
Corresponding PV & $q=\nabla^{2}\psi-\psi+\beta y$ & $q=\widetilde{\Phi}-\nabla^{2}\Phi-\kappa x$\tabularnewline
Characteristic freq. & \makecell{Rossby waves freq. \\ $\omega_{k}=\beta k_{x}/(1+k^{2})$} & \makecell{Drift waves freq. \\ $\omega_{k}=\kappa k_{y}/(1+k^{2})$}\tabularnewline
\hline
\textbf{Minimal unstable model} & \textbf{Two-layer quasi-geostrophic} & \textbf{Hasegawa-Wakatani} \tabularnewline
Instability source & \makecell{Velocity shear $\Delta U$ between \\ the two layers $i=\{1,2\}$} & \makecell{Non-adiabatic electron \\ response due to collisions}\tabularnewline
Perturbed PV & $q^\prime_{1,2}=\nabla^{2}\psi_{1,2}+ F_{1,2}\left(\psi_{2,1}-\psi_{1,2}\right)$ & $q^\prime=n-\nabla^{2}\Phi$\tabularnewline
\hline 
\end{tabular*}\caption{Summary of the analogy between the geophysical and plasma problems. In the geophysical problem, the various parameters are: mean layer height $h_0$, mean Coriolis parameter $f_0$, typical velocity $U$ and scale $L$ of the flow, and $F_i$ is the relative layer height for the two-layer quasi-geostrophic model (see \cite{wang:2016}). For the plasma problem, the parameters are: ion sound speed $c_s$, ion cyclotron frequency $\Omega_{i}$, density inverse gradient length $L^{-1}_n=(1/n)|dn/dx|$, and $\bmE$ and $\bmB$ are the electric and magnetic fields, respectively.\label{table:sum1}}
\end{table}

In this study, we focus mainly on the observation of fundamental differences
in layering between $\beta$-plane like systems that are driven by
instability or by external forcing. In particular, the instability-driven
plasma system is able to self-regulate, not only the turbulence, but also
the energy injection, essentially by the influence of ZFs on the linear
instability that drives the turbulence. External forcing (either local
in configuration space or local in scale, in Fourier space), by contrast,
keeps injecting energy even in a saturated state, and forces the system
to somehow find a way to dissipate the energy/enstrophy that is being
pumped continuously into the system. However, our main observation is that the differences observed in layering is a manifestation of the treatment of zonal modes in each system, or the `zonal response' for short, which is usually different in geophysical and plasma contexts, rather than the nature of energy injection.

For the forced $\beta$-plane case, we consider two different kinds
of zonal flow evolution. In the standard CHM system, the conserved
dimensionless PV is defined as $q=\zeta-\psi+\beta y$, in terms of
the stream function $\psi$ and the vorticity $\zeta\equiv\nabla^{2}\psi$, with $\beta$ the planetary vorticity,
which means that the large-scale zonal PV is proportional to the zonally
averaged stream function (i.e., $\left\langle q\right\rangle _{x}\approx-\left\langle \psi\right\rangle _{x}+\beta y$
as $\partial_{y}\rightarrow0$). On the contrary, in the generalised
Hasegawa-Mima (GHM) system \cite{smolyakov:1999} relevant for fusion
plasmas, the PV is defined as $q=\widetilde{\Phi}-\nabla^{2}\Phi-\kappa x$,
with $\Phi$ being the non-dimensional electrostatic potential (which
plays the role of the stream function) and where $\kappa$ is the mean density gradient, which is the
equivalent of $\beta$ but with the zonal direction being $y$ instead
of $x$. In the plasma case, only the non-zonal part of the stream
function, $\widetilde{\Phi}\equiv\Phi-\langle\Phi\rangle_{y}$, plays
a role in the PV. As a consequence, the large-scale zonal PV for the
GHM is vorticity, instead of the stream function (i.e., $\left\langle q\right\rangle _{y}\approx-\partial^{2}_{x}\langle\Phi\rangle_{y}-\kappa x$
as $\partial_{x}\rightarrow0$). This slightly different form of the
zonal PV is actually a consequence of the zonal flow response. In
the plasma problem, when drift wave perturbations have a finite parallel
gradient along the guiding magnetic field, the electrons move rapidly
in the parallel direction, causing the difference between density
and electrostatic potential to decay. This is called the electron
response. However ZFs, in the plasma problem, vary only in the radial
($x$) direction. This means that, as they have no parallel perturbations,
they are unaffected by the electron response, and hence zonal density
and zonal electrostatic potential remain independent of one another.
This is called the \emph{modified} ZF response, and it causes large-scale
zonal modes to have minimum inertia. As will be shown, the difference
in zonal response has drastic consequences for turbulence self-organisation
properties.

We then consider two examples of instability-driven $\beta$-plane
turbulence models: the Hasegawa-Wakatani model \cite{hasegawa:1983},
which is considered as a minimal model of plasma turbulence and ZFs,
with drift waves driven by a dissipative drift instability, and the
two-layer quasi-geostrophic (QG) model \cite{flierl:1978,wang:2016},
which is a minimal representation of the baroclinic instability. The
former, arising from a plasma problem, has the modified zonal response,
while the latter being a GFD model, incorporates a standard zonal
response (i.e., ZFs are not treated differently from the other modes).
The mechanisms of instability are also very different, since it is
the non-adiabatic (i.e., not infinitely fast) electron response in the parallel direction that
activates the dissipative drift instability, whereas the particular
form of the two-layer model that we consider relies on a vertically
sheared background zonal (east-west) flow that is modelled as a discontinuity
between the two layers.

Nevertheless, in both models, even though the turbulence may be described
using only two dimensions, the instabilities require three-dimensional
effects. For tokamak plasmas \cite{kadomtsev:book:plasturb}, which
are more or less periodic in the third dimension, these can be introduced
by considering a finite wave-number, and a finite electron mobility
in this direction, resulting in the minimal Hasegawa-Wakatani model.
In contrast, in the GFD case \cite{pedlosky-book:71}, the minimal
three dimensionality necessary to describe the baroclinic instability
in planetary atmospheres is introduced by the two-layer QG model,
in which the vertically stacked layers have differing background zonal
velocities as well as densities, resulting in an inverted PV profile.
This underlines once again how the two systems are very different
in details related to geometry, forcing and boundary conditions, while
being very similar in terms of the underlying non-linear structure
of the equations.

The different models examined in the article, their corresponding
forcing, zonal response and observed layering are described in table~\ref{table:models}.
As a short summary of our results, we find that for forced $\beta$-plane
turbulence driven at small-scales by stochastic forcing and with \emph{standard}
zonal response (i.e., CHM driven at small scales), the forcing amplitude
and time scale (i.e., the time step of the random forcing) require
tuning to achieve layering. If the system is forced weakly and sufficiently
slowly, the ZFs form in stages, as suggested by the wave turbulence
picture \cite{balk:90b,nazarenko:2009,connaughton:15}. We note,
however, that if the forcing is too rapid or too strong, 2D Navier-Stokes
like behaviour is instead observed. On the contrary, for the GHM system,
where the zonal response is \emph{modified}, we observe the rapid
formation of straight stationary ZFs, whose size is well defined and
does not seem to change in time, over the time scales of the simulations.
The contrast between the standard zonal response, which generates
layering with low non-zonal modes coming along with purely zonal modes,
and the modified response, which results in a purely zonal pattern,
is illustrated in table~\ref{table:models} by the fraction of zonal
to total kinetic energy, which accounts only for purely zonal modes.

In the two-layer model, which has the standard response, the saturated
state is again dominated by strong elliptic ZFs that are modulated
and seem to play the role of wave-guides for the remaining Rossby
waves \cite{dubrulle:97}. By contrast, in the Hasegawa-Wakatani
system, two major regimes can be reached: either a ZF-dominated regime
or a 2D turbulence regime without flows, depending on the ratio of
two linear parameters, the adiabatic coupling $C$ and the mean density
(i.e., PV) gradient $\kappa$. For $C/\kappa>0.1$, one finds a state
that is dominated by ZFs at large scales, and displays characteristic
features of a two-dimensional forward enstrophy cascade at small scales.
Since we choose not to introduce zonal flow damping unless necessary,
the saturated state has zero net Reynolds power --- i.e., the energy
transfer associated with the Reynolds stress. Thus, once established,
the flows turn off the energy drive, and the system ends up in a fixed-point-like
stationary state, with straight jets having a well-defined spatial
scale. The Hasegawa-Wakatani system is also well known to exhibit
an abrupt transition, with a hysteresis loop, from a zonally dominated
state to a state of two-dimensional turbulence as its linear control
parameter $C/\kappa$ is varied \cite{numata:2007,grander:2024,guillon:2025}.
Note, however, that extending this observation to gyrokinetic systems
where turbulence is driven by the ion temperature gradient instability
or trapped electron modes remains unclear \cite{dimits:2000}.

\begin{table}
\centering{}%
\begin{tabular*}{1\textwidth}{@{\extracolsep{\fill}}@{\extracolsep{\fill}}lcccc}
\hline 
Model & $\beta$-plane/CHM & GHM & Two-layer QG & HW\tabularnewline
\hline
Forcing & \multicolumn{2}{c}{Small scale stochastic} & \multicolumn{2}{c}{Linear instability}\tabularnewline
Equation(s) & (\ref{eq:chm}) & (\ref{eq:ghmend}) & (\ref{eq:2lqga}) \& (\ref{eq:2lqgb}) & (\ref{eq:hwd}) \& (\ref{eq:hwv})\tabularnewline
Zonal response & Standard & Modified & Standard & Modified\tabularnewline
Layering & \makecell{Elliptic jets \\  merging} & \makecell{Straight jets \\ finite size} & \makecell{Elliptic jets \\ modulated} & \makecell{Straight jets \\ finite size}\tabularnewline
\makecell[l]{Zonal to total \\ kinetic energy ratio} & $36.4 \pm 1.0 \%$ & $97.3 \pm 0.3 \%$ & $24.3 \pm 0.9 \%$ & $91.5 \pm 0.3 \%$ \tabularnewline
\hline 
\end{tabular*}\caption{The different models examined in the article, their corresponding
forcing, zonal response and observed layering, with the fraction of zonal to total kinetic energy averaged over the last quarter of each simulation.}
\label{table:models}
\end{table}

The rest of the paper is organised as follows. In \S\,\ref{sec:Models}
we introduce the different models based on PV conservation, and their
corresponding minimal models with instability, which drive turbulence
and then layering. In \S\,\ref{sec:Forced-beta-plane} we examine
the layering process in $\beta$-plane turbulence (\emph{via} CHM,
two-layer QG and GHM models), which can be forced either randomly
at small scales, or coherently at large scale \emph{via} a linear
instability mechanism. In \S\,\ref{sec:HWFG} we detail the instability-driven
plasma case, discussing the formation of ZFs in the HW system, which
have a finite size without large-scale friction and which turn off
the linear drive.

\section{Models for geophysical and plasma turbulence}

\label{sec:Models}

In this section, we define the four different models based on $\beta$-plane
turbulence that exhibit layering at the most simple level. First,
in \S\,\ref{sec:Models}\ref{sec:betaplane}, we start from PV conservation
to derive the standard Charney-Hasegawa-Mima (CHM) model, and its
equivalent in the plasma context, which has the modified zonal response
--- the Generalised Hasegawa-Mima (GHM) system. Both of these models
have to be forced at small scales in order to generate 2D turbulence
that then develops layering. We then go on to discuss how the responses
can affect the layering.

Then, in \S\,\ref{sec:Models}\ref{sec:minimal}, we consider the
minimal models in the $\beta$-plane framework where the energy is
injected self-consistently through a linear instability: the two-layer quasi-geostrophic (QG) model in the GFD case (which
has the standard zonal response), and the Hasegawa-Wakatani (HW)
model in the plasma context (which has the modified zonal response).

\subsection{PV conservation and $\beta$-plane turbulence}

\label{sec:betaplane}

\subsubsection{Geophysical problem}

For a fluid in large scale equilibrium around a planet, such as the
atmosphere or ocean, gravity $g\bmzhat$ is balanced by the vertical
pressure gradient $\partial_{z}P$, in what is known as hydrostatic
balance. This naturally links the fluid pressure variations to variations
in height through a barometric relation such as $\Delta P=-\rho g\Delta h$, with $\rho$ the fluid density.
The basic equation for a rotating fluid in such a vertical hydrostatic
balance is the Navier-Stokes equation, written in a rotating frame:
\begin{equation}
\left(\partial_{t}+\bmu\cdot\bmnabla\right)\bmu+2\bmOmega_{p}\times\bmu=-\frac{\bmnabla P}{\rho}-g\bmzhat+\nu\nabla^{2}\bmu,\label{eq:ns}
\end{equation}
where the last term on the left hand side represents the Coriolis
force due to planetary rotation $\bmOmega_{p}$. While the hydrostatic
balance defines a direction $\bmnhat$ normal to the atmosphere, which
can be used to naturally decompose the motion of the fluid into parallel
and perpendicular directions, the background rotation $\bmOmega_{p}$
defines a second axis. Considering the projection of the planetary
vorticity implied by the rotation of the planet onto the direction
normal to the atmosphere, one defines the Coriolis parameter, or planetary vorticity, as
\begin{equation}
f\equiv2\bmOmega_{p}\cdot\bmnhat=2\Omega_{p}\sin\theta,\label{eq:f}
\end{equation}
where $\theta$ is the latitude. The $\beta$-plane can then be defined
as a tangential plane at a (central) latitude $\theta$, with the
local coordinates $x$ and $y$ corresponding to zonal (east) and
meridional (north) directions (see figure~\ref{fig:Geometries} for
an illustration).

For strong rotation (characterised by small Rossby number $Ro$, where
$Ro$ is the ratio of inertial to Coriolis forces), the principal
(geostrophic) balance is between the Coriolis force and the horizontal
pressure gradient; this represents a balance between fluctuating fields,
in contrast to the hydrostatic balance between equilibrium quantities.
Geostrophic balance, together with the barometric relation between
pressure and height, can be used to write \cite{vallis:book} 
\begin{equation}
\bmu_{\perp}=\frac{\bmnhat\times\bmnabla P}{f\rho}=\frac{g}{f}\bmnhat\times\bmnabla h,\label{eq:geob}
\end{equation}
thus implying that the height $h$ plays the role of the stream function
$\psi$ (i.e., as in $\bmu=\bmnhat\times\bmnabla\psi$). Note that
this relation can formally be obtained as the leading order term in
a small Rossby number expansion of the Navier-Stokes equation.

The key aspect of the $\beta$-plane model is the assumption that
Eqn.~(\ref{eq:f}) can be expanded in a Taylor series about the central
latitude $\theta$, as $f=f_{0}+\beta y$, with $\beta=\drm f/\drm y|_{\theta}$.
On defining the relative vorticity in the normal direction as $\zeta=\bmnhat\cdot\left(\bmnabla\times\ \bmu\right)$,
and using the continuity equation for the layer thickness, one can
show that 
\begin{equation}
\frac{\drm}{\drm t}\left(\frac{\zeta+f}{h}\right)=0.\label{eq:pvc}
\end{equation}
Eqn.~\eqref{eq:pvc} represents the conservation of PV, $q=(\zeta+f)/h$,
implying that the total vorticity (fluid $+$ planetary) and the height
of the fluid increase or decrease together.

On writing $h=h_{0}+h_{1}$, together with $f=f_{0}+\beta y$, with
$h_{1}/h_{0}$ and $\beta y/f_{0}$ coming into play at the same order
in the small Rossby number expansion (e.g., see \cite{vallis:book}),
the PV takes the form 
\begin{equation}
q=\frac{\zeta+f_{0}+\beta y}{h_{0}+h_{1}}\approx\frac{\zeta+f_{0}+\beta y-f_{0}h_{1}/h_{0}}{h_{0}}.\label{eq:pvd}
\end{equation}

Using (\ref{eq:geob}) with this expansion to define $\psi\equiv\left(g/f_{0}\right)h_{1}$,
and removing the constant $f_{0}/h_{0}$, yields 
\begin{equation}
q\approx h^{-1}_{0}\left(\nabla^{2}\psi+\beta y-R^{-2}\psi\right),\label{eq:pvdef}
\end{equation}
with $R\equiv\sqrt{gh_{0}}/f_{0}$, the Rossby deformation radius,
defining a characteristic length scale and $f^{-1}_{0}$ a characteristic
time scale. On defining $x/R\rightarrow x$, $f_{0}t\rightarrow t$,
$\psi/(f_{0}R^{2})\rightarrow\psi$, and $R\beta/f_{0}\rightarrow\beta$
as dimensionless variables, and rescaling the PV by $h_{0}$, the
non-dimensional form of the Charney-Hasegawa-Mima (CHM) equation can
be written as 
\begin{equation}
\frac{\drm q}{\drm t}=\partial_{t}\left(\nabla^{2}\psi-\psi\right)+\beta\partial_{x}\psi=-(\bmnhat\times\bmnabla\psi)\cdot\bmnabla(\nabla^{2}\psi),\label{eq:chm}
\end{equation}
where $q=\nabla^{2}\psi-\psi+\beta y$ is the potential vorticity.
By adding forcing and dissipation, we obtain the forced $\beta$-plane
model, 
\begin{equation}
\left(\partial_{t}+(\bmnhat\times\bmnabla\psi)\cdot\bmnabla\right)q=\mathsf{f}_{q}+\nu\nabla^{2}q,\label{eq:fbp}
\end{equation}
where $\mathsf{f}_{q}$ is the forcing and $\nu\nabla^{2}q$ is the
dissipation. Note that, here, we consider the case of a finite Rossby deformation radius R, but it is also common to explore the large $R$ limit by
replacing the fluctuating part of $q$ by $\zeta$.

\subsubsection{Plasma problem}

In contrast to hydrostatic balance in the geophysical problem, the
large-scale balance in the plasma problem is represented by the MHD
equilibrium between the Lorentz ($\bmJ\times\bmB$) force, due to
equilibrium magnetic fields $\bmB$ and currents $\bmJ$, and the background pressure
gradient. This equilibrium imposes a direction $\bmbhat$ in the direction
of the equilibrium magnetic field, which can be used to separate the
motion parallel and perpendicular to the magnetic field. In the plasma
framework, the local coordinate $x$ corresponds to the radial (from
tokamak core to the edge) and $y$ to the poloidal, or bi-normal direction
(i.e., $\bmyhat=\bmbhat\times\bmxhat$ is roughly poloidal). The toroidal
direction, parallel to the magnetic field (i.e., to $\bmbhat$), is denoted by $z$ (see
figure~\ref{fig:Geometries} for an illustration). Contrary to the
GFD case, the PV gradient here is along $x$ and zonal flows are uniform
along the $y$ direction.

Considering ions as a fluid, the ion equation of motion takes the
form 
\begin{equation}
\left(\partial_{t}+\bmu_{i}\cdot\bmnabla\right)\bmu_{i}=-\frac{\bmnabla P}{m_{i}n_{i}}+\frac{e}{m_{i}}\left(\bmE+\bmu_{i}\times\bmB\right),\label{eq:ieom}
\end{equation}
where $e$ is the unit charge, $n_i$ and $m_i$ are the ion density and mass, respectively, and the ion velocity $\bmu_{i}$ is a sum of drift velocities. We consider here the electrostatic case, so that the electric field is $\bmE=-\nabla\Phi$, where $\Phi$ is the electrostatic potential. The
MHD stationary equilibrium follows by considering ions and electrons
together as a single fluid in static equilibrium. Using a formal drift
expansion in a strong magnetic field (i.e., with the ratio of the
inverse characteristic time scale to the cyclotron frequency being
the small parameter), the leading order term is the $\bmE\times\bmB$
drift velocity, which can be written as 
\begin{equation}
\bmu_{E}=\frac{\bmbhat\times\nabla\Phi}{B}=\bmbhat\times\bmnabla\left(\frac{e\Phi}{T_e}\right)\rho_{s}c_{s},\label{eq:ExB}
\end{equation}
where the sound speed is $c_{s}=\sqrt{T_{e}/m_{i}}$, with $T_e$ the electron temperature, the sound Larmor
radius is $\rho_{s}=c_{s}/\Omega_{i}$ and $\Omega_{i}=eB/m_{i}$ is the
ion cyclotron frequency. The last two quantities define respectively
the spatial and temporal scales of the problem. The plasma, considered
as a fluid, moves with the $\bmE\times\bmB$ velocity, since both
ions and electrons move together at that speed in the same direction.
Comparing (\ref{eq:ExB}) with (\ref{eq:geob}), we see that the geostrophic
balance and the $\bmE\times\bmB$ drift velocity $\bmu_{E}$ have
the same structure. The stream function role in the plasma problem
is now played by the electrostatic potential $\Phi$.

To derive the $\beta$-plane equivalent in the plasma context, one
usually starts by considering the equation of continuity (see, e.g.,
\cite{hasegawa:1978,horton:94} for a complete derivation), 
\begin{equation}
\left(\partial_{t}+\bmu_{E}\cdot\bmnabla\right)n=\frac{1}{e}\nabla_{\parallel}J_{\|},\label{eq:contne}
\end{equation}
where the plasma density is $n=n_{e}=n_{i}$ due to quasi neutrality,
and $J_{\|}$
corresponds to the electron current in the direction parallel to $\bm{B}$.
Here we assume a uniform and constant magnetic field $\bm{B}$, and
constant electron temperature $T_{e}$, so that the divergence of
the particle flux reduces to $\bmu_{E}\cdot\bmnabla n$. Then, charge
conservation, together with the assumption of cold ions (i.e., $T_{i}=0$),
and keeping the first term in the small parameter $\rho_{s}/L_{n}$
expansion, where $L_{n}$ is the characteristic length of the density
gradient, leads to the vorticity equation, 
\begin{equation}
-n\rho^{2}_{s}\left(\partial_{t}+\bmu_{E}\cdot\bmnabla\right)\nabla^{2}\left(\frac{e\Phi}{T_{e}}\right)=\frac{1}{e}\nabla_{\parallel}J_{\parallel},\label{eq:chargcons}
\end{equation}
where the left hand side accounts for the divergence of the so-called
ion polarisation current.

To further simplify the system, one uses the parallel component of
Ohm's law without collisions, which gives $\nabla_{\parallel}P=en\nabla_{\parallel}\Phi$,
thus relating parallel variations of pressure $P$ and electrostatic
potential $\Phi$. In combination with the ideal gas equation of state
$P=nT$, with $T$ understood as thermal energy, we obtain the so-called
\textit{adiabatic} electron response: 
\begin{equation}
\frac{\widetilde{n}}{n_{0}}\approx\frac{e\widetilde{\Phi}}{T}\;\text{,}\label{eq:ade}
\end{equation}
which applies only on the \textit{non-zonal} component of the field $A$, defined as $\widetilde{A}\equiv A -\langle A\rangle_y$. The origin of this particular form is detailed later.

In the plasma problem, the break down of the adiabatic electron response
gives rise to various drift instabilities. Even though there exists
a variety of different mechanisms leading to the break down of adiabaticity,
resulting in different kinds of drift instabilities, here, in \S\,\ref{sec:Models}\ref{sec:minimal},
we consider only the effect of collisions in the parallel direction,
as a minimal example of the dissipative drift instability. But first,
we assume that the adiabatic electron response (\ref{eq:ade}) holds,
so that we obtain the equivalent of the $\beta$-plane model in the
plasma problem.

The fact that the adiabatic electron response (\ref{eq:ade}) applies
only for non-zonal fields, i.e., $\widetilde{\Phi}=\Phi-\langle\Phi\rangle_{y}$
and $\widetilde{n}=n-\langle n\rangle_{y}$, underlines a non-trivial
feature of the plasma geometry. In the plasma problem, ZFs are constant
on magnetic flux surfaces (i.e., along $y$ and in the $\bmbhat$ -- or parallel --
direction), and they vary only in the radial (i.e., $x$ in the plasma
convention) direction. Hence, while the non-zonal modes, which are
denoted by $\widetilde{\Phi}$, $\widetilde{n}$ above, vary in all three directions
and their variation in the parallel direction is governed by the adiabatic
electron response, zonal perturbations are unaffected by this response.
In a sense, electrons cannot respond to (zonal) perturbations that
they do not see (their small gyro-radius does not allow them to feel
radial-only perturbations). As such, zonal electrostatic potential
and zonal density behave independently of each other. This is the
physical basis of the \emph{modified} zonal flow response for the
plasma problem; as a consequence, the PV definition is slightly modified,
as shown below.

To derive the $\beta$-plane plasma model, we take the zonal density
$\overline{n}\equiv\langle n\rangle_{y}=0$. We can thus expand $n=n_{0}(-x/L_{n}+\widetilde{n})$,
taking a mean profile linearly decreasing as a function of radius,
while $\Phi=\overline{\Phi}+\widetilde{\Phi}$, where $\overline{\Phi}\equiv\langle\Phi\rangle_{y}$
is the zonal electrostatic potential. Using the adiabatic electron
response, and combining Eqns.~\eqref{eq:contne} and \eqref{eq:chargcons},
we obtain the so-called generalised-Hasegawa-Mima equation \cite{smolyakov:1999,smolyakov:00},
which can be written in non-dimensional variables $e\Phi/T\rightarrow\Phi$,
$x/\rho_{s}\rightarrow x$, $y/\rho_{s}\rightarrow y$ and $\Omega_{i}t\rightarrow t$
as 
\begin{equation}
\frac{\drm q}{\drm t}=\left(\partial_{t}+\bmbhat\times\bmnabla\Phi\cdot\bmnabla\right)\left(\widetilde{\Phi}-\nabla^{2}\Phi\right)+\kappa\partial_{y}\Phi=0,\label{eq:ghmend}
\end{equation}
where $\kappa\equiv-\rho_{s}\frac{1}{n_{0}}\frac{dn_{0}}{dx}=\rho_{s}/L_{n}$
is the mean density gradient. In this case, the conserved PV has the
form $q=\widetilde{\Phi}-\nabla^{2}\Phi-\kappa x$, with the common
sign convention in the plasma community. Note that here, $\rho_{s}/L_n$
is usually a small parameter and therefore the vorticity term is a
small correction to the density term at large scales.

\subsubsection{Standard and modified zonal flow responses}

\label{sec:resp} Considering their 3D structure, in the GFD problem,
zonal perturbations are invariant in the $x$ (i.e., east-west) direction,
so they vary only in the $y$ (i.e., north-south) direction, but there
is nothing special about their variation in the direction normal to
the atmosphere; they vary together with the other modes through the
relation to fluid height. Hence, the zonal PV is $\langle q\rangle_{x}=\partial^{2}_{y}\langle\psi\rangle_{x}-\langle\psi\rangle_{x}+\beta y$,
and, consequently, the time evolution of the zonal component of the
stream function $\overline{\psi}\equiv\langle\psi\rangle_{x}$ in
Fourier space can be written from Eqn.~\eqref{eq:chm} as 
\begin{equation}
\partial_{t}\overline{\psi}_{k}=\frac{\mathcal{F}_{k}[\langle\hat{\mathbf{n}}\times\bmnabla\psi\cdot\bmnabla\nabla^{2}\psi\rangle_{x}]}{1+k^{2}},\label{eq:zfrespgfd}
\end{equation}
where $\overline{\psi}_{k}$ is the Fourier component of $\overline{\psi}$
with zonal wave vector $\bm{k}=(0,k_{y})$, and $\mathcal{F}_{k}$
is the Fourier transform operator to the $k$ mode. This corresponds
to the \emph{standard} zonal response.

On the contrary, in the plasma problem, as previously discussed, the
zonal PV is modified owing to the adiabatic electron response~\eqref{eq:ade}
applying only to non-zonal perturbations, and becomes $\langle q\rangle_{y}=-\partial^{2}_{x}\langle\Phi\rangle_{y}-\kappa x$.
As such, using Eqn. \eqref{eq:ghmend}, the time evolution of the zonal
Fourier component $\overline{\Phi}_{k}$ of the electrostatic potential
(where now $\bm{k}=(k_{x},0)$ in the plasma problem coordinates)
is described by 
\begin{equation}
\partial_{t}\overline{\Phi}_{k}=-\frac{\mathcal{F}_{k}[\langle\hat{\mathbf{b}}\times\bmnabla\Phi\cdot\bmnabla\nabla^{2}\Phi\rangle_{y}]}{k^{2}},\label{eq:zfrespla}
\end{equation}
which constitutes the \emph{modified} zonal response. The main difference
with the GFD case (\ref{eq:zfrespgfd}) is that for large-scale ZFs,
i.e., $k\to0$, the right hand side of Eqn. \eqref{eq:zfrespla} grows
as $1/k^{2}$, whereas it remains bounded for the standard zonal response
(i.e., $1/(1+k^{2})\to1$). Therefore, it is said that ZFs have more
\emph{inertia} in standard $\beta$-plane turbulence, as opposed to
the plasma case with the modified response, where they are much more
easily excited, especially at large scales, which can explain why
the flows are straighter and more easily formed. Note that in the limit of an infinite Rossby deformation radius, the geophysical PV (\ref{eq:pvdef}) in the CHM equation becomes $q=\nabla^2\psi+\beta y$, and as a consequence ZFs loose their inertia since their response becomes similar to the modified one. In that case, the observed jets are also straighter. 

\subsection{Minimal models with instability-driven turbulence}

\label{sec:minimal} Both CHM and GHM models allow for the propagation
of waves, respectively Rossby and drift waves, which are linearly
stable. Therefore, to generate turbulence and consequently layering,
one has to force the systems --- e.g., stochastically at small scales.

Another mechanism for energy injection, which is physically more relevant,
especially for the plasma case, is via a linear instability. In the following,
we consider the minimal model based on $\beta$-plane turbulence featuring
a linear instability: the two-layer quasi geostrophic model for the
geophysical case, which has the standard zonal response, and the Hasegawa-Wakatani
system for the plasma problem, which has the modified zonal response.

\subsubsection{Geophysical case: the two-layer QG model}

There are a number of similarly `simplest' models in GFD that may
be used to describe the baroclinic instability. In these models, it
is the vertical flow shear that gives rise to the basic linear instability,
which can be modelled by considering two layers moving at different
speeds. The model that we study here consists of PV conservation equations
in each layer \cite{flierl:1978,wang:2016}: 
\begin{equation}
\left(\partial_{t}+U_{1}\partial_{x}+\bmnhat\times\bmnabla\psi_{1}\cdot\bmnabla\right)q^\prime_{1}+\left(\beta+F_{1}\Delta U\right)\partial_{x}\psi_{1}=0,\label{eq:2lqga}
\end{equation}
\begin{equation}
\left(\partial_{t}+U_{2}\partial_{x}+\bmnhat\times\bmnabla\psi_{2}\cdot\bmnabla\right)q^\prime_{2}+\left(\beta-F_{2}\Delta U\right)\partial_{x}\psi_{2}=0,\label{eq:2lqgb}
\end{equation}
where $U_{i}$ is the east-west velocity of the $i$-th layer, such
that $\Delta U=U_{1}-U_{2}$ makes the two layers vertically sheared.
In this system, the total PV of the $i$-th layer, $q_{i}$, is 
\begin{equation}
q_{1}=q^{\prime}_{1}+\left(\beta+F_{1}\Delta U\right)y\hspace{35pt}\hbox{and}\hspace{35pt}q_{2}=q^{\prime}_{2}+\left(\beta-F_{2}\Delta U\right)y,
\end{equation}
where the perturbed PV (zonal and non-zonal combined) $q^{\prime}_{i}$
is 
\[
q^{\prime}_{1}=\nabla^{2}\psi_{1}+F_{1}\left(\psi_{2}-\psi_{1}\right)\hspace{35pt}\hbox{and}\hspace{35pt}q^{\prime}_{2}=\nabla^{2}\psi_{2}+F_{2}\left(\psi_{1}-\psi_{2}\right),
\]
with $i=\{1,2\}$ corresponding to the top and bottom layers respectively.
Here, $F_i$ is the relative height of the layer $i$ (see \cite{wang:2016}), and $\beta+F_{1}\Delta U$ and $\beta-F_{2}\Delta U$ represent
the background PV gradients, due to planetary vorticity and the vertically
sheared background (east-west) mean flows. Using the normalisation
$x/R\rightarrow x$, $y/R\rightarrow y$ , $Ut/R\rightarrow t$, $R^{2}\beta/U\rightarrow\beta$,
$F_{i}/R^{2}\rightarrow F_{i}$ and finally $\psi_{i}R^{-1}U^{-1}\rightarrow\psi_{i}$,
so that $q_{i}R/U\rightarrow q_{i}$, and physical parameters from
\cite{wang:2016}, the linear growth rate peaks around the dimensionless
wave-number $k\approx0.6$. Since, for these parameters, $\beta$
is rather small compared to $U/R^{2}$, the Rossby wave-number $k_{R}=\sqrt{\beta/U}$
computed with the imposed jet speed $U$ is smaller than the wavenumber
of the most unstable mode. Thus the box size should be sufficiently
large in order to capture the inverse energy cascade, which produces
ZFs at large enough scales, where the Rossby wave dynamics can take
over. Bottom friction can be included in the equation for $q_{2}$
[i.e., a term of the form $\nu_{f}\nabla^{2}\psi_{2}$ on the left
hand side of Eqn.~\eqref{eq:2lqgb}], as well as viscous dissipation
in both equations [i.e., $\nu\nabla^{2}q_{1,2}$ on the right hand
sides of Eqns.~\eqref{eq:2lqga} and \eqref{eq:2lqgb}] to remove
small-scale potential enstrophy.

\subsubsection{Plasma case: the Hasegawa-Wakatani model}

As we mentioned in the previous section, breaking of the adiabatic
electron response in the parallel direction (\ref{eq:ade}) promotes
drift-wave instabilities; one of the simplest ways this can happen
is through finite electron mobility --- or non-zero resistivity $\eta$.
In this case, Ohm's law in the parallel direction yields a parallel
current 
\begin{equation}
J_{\parallel}=\frac{T_e}{e\eta}\nabla_{\parallel}\left(\frac{e}{T_e}\widetilde{\Phi}-\frac{\widetilde{n}}{n_{0}}\right),
\label{eq:ohmsfinite}
\end{equation}
which we can insert into the right hand sides of Eqns. (\ref{eq:contne})
and (\ref{eq:chargcons}).

We can then write the density equation in dimensionless variables
as before, now using $n_{0}(-\kappa x+\overline{n}\left(x,t\right)+\widetilde{n}\left(x,y,t\right))$
as total density: 
\begin{equation}
\left(\partial_{t}+\bmbhat\times\bmnabla\Phi\cdot\bmnabla\right)n+\kappa\partial_{y}\widetilde{\Phi}=C\left(\widetilde{\Phi}-\widetilde{n}\right),\label{eq:hwd}
\end{equation}
where we have assumed a single dominant parallel wave-number $k_{\parallel}$ (e.g., $\widetilde{\Phi}\left(x,y,z\right)\propto\widetilde{\Phi}\left(x,y\right)\cos k_{\parallel}z$)
in order to define $C\equiv v^{2}_{te}k^{2}_{\parallel}/\left(\nu_{c}\Omega_{i}\right)$
as a model parameter, where $\nu_{c}$ is the collision frequency
(which sets the resistivity $\eta=m_{e}\nu_{c}/e^{2}n_{0}$) and $v_{te}$
is the electron thermal velocity. Similarly, the non-dimensionalised
vorticity equation becomes 
\begin{equation}
\left(\partial_{t}+\bmbhat\times\bmnabla\Phi\cdot\bmnabla\right)\zeta=C\left(\widetilde{\Phi}-\widetilde{n}\right),\label{eq:hwv}
\end{equation}
where $\zeta=\nabla^{2}\Phi$ is the vorticity. Equations~\eqref{eq:hwd}
and \eqref{eq:hwv} together constitute the inviscid version of the
Hasegawa-Wakatani system. Subtracting Eqn.~\eqref{eq:hwv} from \eqref{eq:hwd}
yields the equation of PV conservation, 
\begin{equation}
\frac{\drm q}{\drm t}=\left(\partial_{t}+\bmbhat\times\bmnabla\Phi\cdot\bmnabla\right)q^{\prime}+\kappa\partial_{y}\Phi=0,\label{eq:hwq}
\end{equation}
where $q=q^{\prime}-\kappa x$ with $q^{\prime}=n-\zeta$ the perturbed
PV. In this case, the modified zonal response manifests through the
coupling term $C(\widetilde{\Phi}-\widetilde{n})$, where $C$ is
known as the \emph{adiabaticity} parameter. Considering the linearised
time evolution of $(\widetilde{n}-\widetilde{\Phi})$ in Fourier space,
we obtain 
\begin{equation}
\partial_{t}\left(\widetilde{n}_{k}-\widetilde{\Phi}_{k}\right)+i\kappa k_{y}\widetilde{\Phi}_{k}=-C\left(1+\frac{1}{k^{2}}\right)\left(\widetilde{n}_{k}-\widetilde{\Phi}_{k}\right),\label{eq:zfrespHW}
\end{equation}
where $k$ is a non-zonal wave-number (i.e., with $k_{y}\neq0$),
which means that the electron adiabatic response (\ref{eq:ade}) is
recovered at a rate $C(1+1/k^{2})$ for non-zonal perturbations, while
zonal density and electrostatic potential can still behave independently,
as already discussed for the GHM system. Note that, historically,
the HW system was derived without proper attention to the zonal response,
i.e., with $C(\Phi-n)$ rather than $C(\widetilde{\Phi}-\widetilde{n})$
on the right hand sides of (\ref{eq:hwd}) and (\ref{eq:hwv}). Consequently,
in the original system \cite{hasegawa:1983}, ZFs were observed only
for very large values of $C$ \cite{pushkarev:13}. Because of this,
the current form of the HW system is sometimes referred to as the
\emph{modified} Hasegawa-Wakatani model. However, we think that this
is the only correct form of the model, and therefore we do not use
that terminology in this paper.

We present the HW model here as the minimal model of plasma turbulence
because it: i) conserves PV, ii) is driven by an instability whose
free energy is the PV gradient, iii) generates zonal flows, iv) saturates
by a combination of ZFs and the forward enstrophy cascade, and v)
does not seem to result in a condensate state (i.e., the ZF size,
which stays constant, is smaller than the largest available scale)
even without any large scale dissipation.

In order to solve this system numerically, we need to introduce small
scale dissipation terms. Here we use standard diffusion, i.e. $\nu\nabla^{2}\zeta$
and $D\nabla^{2}n$, with the coefficients set to have the smallest
value that results in a saturated state with the given resolution
(details are given in \cite{guillon:2025}).

Another reason we focus on such a simple model is that we can solve
for the linear eigenmode problem analytically. In particular, taking
the Fourier transforms of the linearised forms of Eqns.~(\ref{eq:hwd})
and (\ref{eq:hwv}) and solving for the resulting linear dispersion
relation, we find that the complex eigenfrequency can be written as
\begin{equation}
\omega^{\pm}_{k}=\omega^{\pm}_{kr}+i\gamma^{\pm}_{k}=\pm W_{k}-iA_{k},\label{eq:freq}
\end{equation}
with $\omega^{\pm}_{kr}$ the real frequency and $\gamma^{\pm}_{k}$
the growth/damping rate, where, in the inviscid limit $\nu=D=0$, $A_{k}=C(1+k^{2})/(2k^{2})$ and 
\begin{equation}
W_{k}=\frac{A_{k}}{\sqrt{2}}\left(\frac{k_{y}}{\left|k_{y}\right|}\sqrt{\sqrt{1+\left(2\frac{\omega^{HM}_{k}}{A_{k}}\right)^{2}}-1}+i\sqrt{\sqrt{1+\left(2\frac{\omega^{HM}_{k}}{A_{k}}\right)^{2}}+1}\right),
\end{equation}
where $\omega^{HM}_{k}=\kappa k_{y}/(1+k^{2})$ is the Hasegawa-Mima
frequency, and $k^{2}=k^{2}_{x}+k^{2}_{y}$.

Note that, if we take the limit $C\to \infty$, which physically corresponds to the zero resistivity limit $\eta\to 0$, we recover, from Eqn.~(\ref{eq:ohmsfinite}), the adiabatic response for non-zonal modes (for which the parallel divergence is non-zero), i.e. $\widetilde{\Phi}=\widetilde{n}$. Injecting this relation in Eqn.~(\ref{eq:hwq}) and considering the limit $\overline{n}\to 0$, we recover the GHM equation \eqref{eq:ghmend} and the PV expression $q=\widetilde{\Phi}-\nabla^{2}\Phi-\kappa x$. For $C\gg1$, the HW system converges to the GHM equation at leading order, since relaxation to adiabaticity becomes very strong, while the energy
injection becomes vanishingly small. We can show that the growth rate
in this $C\to+\infty$ limit scales as $\gamma_{k}=k^{2}(\omega^{HM}_{k})^{2}/C(1+k^{2})$,
and peaks at $k_{y}=\sqrt{2}$, rather close to the eventual peak
of the $k$-spectrum, which is dominated by ZFs. This is in contrast
to small-scale injection in the GFD case.

Having described the different minimal models that feature the formation
of ZFs in both GFD and magnetised plasma physics, in the following
section we describe the results of numerical simulations performed
to investigate the layering process in each model, which depends on
the energy injection scale and zonal response.

\section{Forced $\beta$-plane turbulence}

\label{sec:Forced-beta-plane}

The definition of PV (i.e., $q$) in terms of various fields of the
system, and its inversion, defines the instability mechanism, while
PV conservation provides the physics of flow self-organisation. Using
a linear relation between the other fields and the potential (e.g.,
$n_{k}=\left(1+i\delta_{k}\right)\Phi_{k}$ for the Hasegawa-Wakatani
system), one obtains a linear inversion rule, which, when substituted
into the linearised PV conservation equation, gives the dispersion
relation. Using linear inversion, together with the solution of the
linear dispersion relation, the PV conservation equation can be written
with a (small) linear growth term, $\gamma_{k}$ say, representing
the effect of the underlying linear instability pumping PV. If $\gamma_{k}$
is non-zero only at very small scales and we are interested in the
large-scale evolution, we could model the instability as small-scale
stochastic forcing and use Eqn.~\eqref{eq:fbp} invoking an inverse
cascade argument.

In this section, we study the formation of ZFs in the general $\beta$-plane
framework. First, we consider the classic CHM system, forced at small
scales, and then the minimal GFD system with instability, i.e., the
QG two-layer model. We then turn to magnetised plasmas with modified
zonal response: the GHM system forced at small scales, and the Hasegawa-Wakatani
system, the latter is studied in the next section. Animations showing the time evolution of each system can be found
on the repository page \cite{guillon:2026b}.

In the following, we study the energy spectrum of each system. The
kinetic energy $e_{k}(k_{x},k_{y})$ at the wave-number $k$ is defined
as $e_{k}=k^{2}|\psi_{k}|^{2}$ in the geophysical case, and $e_{k}=k^{2}|\Phi_{k}|^{2}$
in the plasma case. The isotropic kinetic energy spectrum $E(k)$
for a given wave-number amplitude $k=\sqrt{k^{2}_{x}+k^{2}_{y}}$
can then be obtained by summing over all the modes inside a ring
of radius $k$ and width $\Delta k$ (taken here as the Fourier grid
increment). The zonal energy spectrum corresponds to the kinetic energy
of the zonal modes, i.e. $E_{Z}(k)=e_{k}(k_{x}=0,k_{y})$ in the geophysical
case, or $E_{Z}(k)=e_{k}(k_{x},k_{y}=0)$ in the plasma case. Finally,
one can also define the non-zonal, turbulent spectrum as $E_{turb}=E-E_{Z}$.

\subsection{Standard Charney-Hasegawa-Mima model forced at small scales}

\begin{figure}[htbp]
\centering{}\includegraphics[width=1\textwidth]{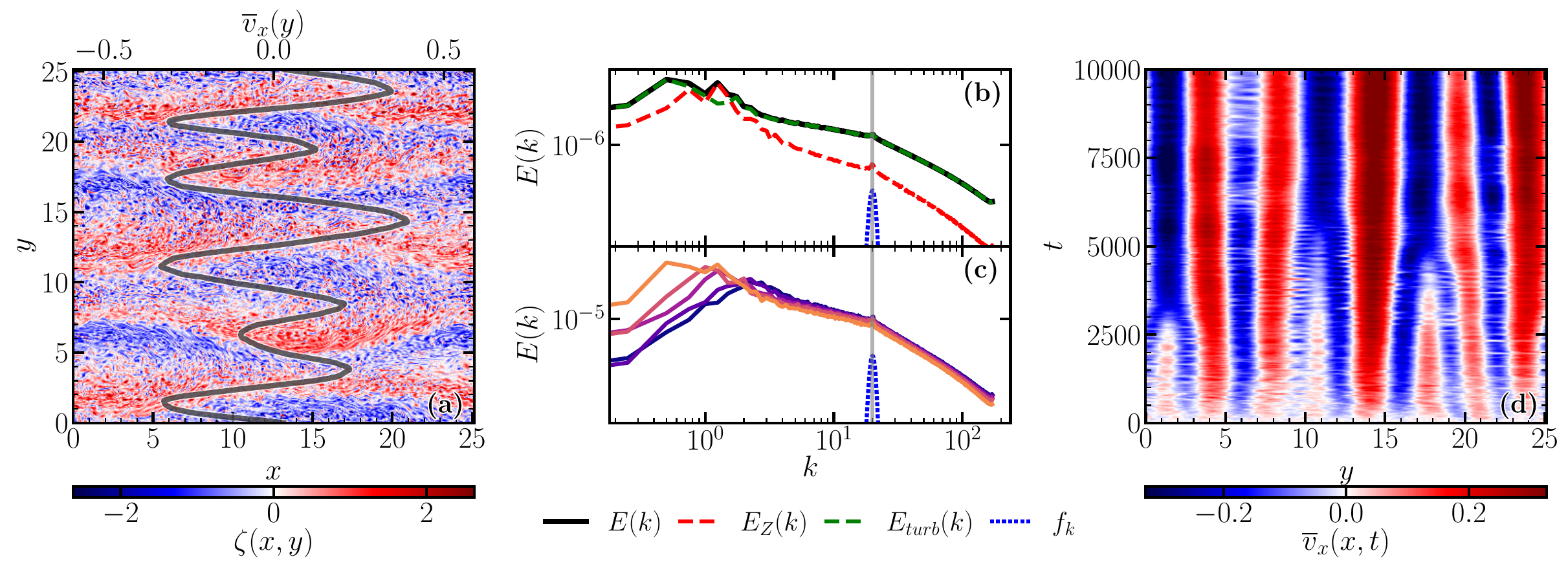}\caption{Standard CHM simulation with small-scale forcing. (a) Vorticity snapshot
$\zeta(x,y)$ at the final time step and the zonal velocity profile
$\overline{v}_{x}$ (grey line). (b) Kinetic energy spectrum $E(k)$
(black), time-averaged over $7500<t<10000$, decomposed into zonal
$E_{Z}(k)$ (red) and non-zonal $E_{turb}(k)$ (green) parts. The
forcing around $k=20$ (vertical grey line) is in dotted blue. (c)
Kinetic energy spectrum at different time-steps, from early stage
(blue) to ZFs (yellow). (d) Hovmöller diagram of the zonal velocity
profile $\overline{v}_{x}$.}
\label{fig:HMsmallforced} 
\end{figure}

Considering the stochastically forced CHM system (\ref{eq:fbp}),
we perform a $2048^{2}$ padded resolution DNS with a box size $L_{x}\times L_{y}=8\pi\times8\pi$,
$\beta=1$, and standard kinematic viscosity with a coefficient $\nu=2\cdot10^{-5}$.
The forcing $f_{k}$ is taken as a Gaussian ring centred at $k=20$,
an amplitude $A=10$, a width $\sigma_{k}=0.5$ and a random phase
that changes at a time scale that is five times the shortest wave
period (so that the forcing stays coherent on the time scale of the
fastest frequency). The simulation is initialised with a Gaussian
seed in Fourier space, also with random phases, and the system is
evolved up to $t=10^{4}$. The results are shown in figure~\ref{fig:HMsmallforced}.
A time animation of the system can also be found on the repository
page \cite[Movie 1]{guillon:2026b}.

Figure~\ref{fig:HMsmallforced}a shows a vorticity snapshot $\zeta(x,y)$
at the final time-step. We can distinguish the self-organised large-scale
horizontal zonal flow pattern, coexisting with smaller-scale 2D vortices.
The jets are not exactly straight, but are elongated in $y$; this
feature may be related to exact solutions of the 2D Euler equation
based on Jacobi elliptic functions \cite{haslam:12,gurcan:23b}. The properties of the interface between the vorticity structures corresponding to the jets have been studied in \cite{sahoo:2025} in the case of infinite Rossby deformation radius. Note that in the latter case, the jets are more straight (see also \cite{scott:2012}), since the zonal response, in this limit, ends up being very similar to the modified one.
The form of the jets is further illustrated in figure~\ref{fig:HMsmallforced}b,
which shows the isotropic spectrum of the kinetic energy $\mathcal{K}=\sum_{k}k^{2}|\psi_{k}|^{2}$
decomposed into zonal ($k_{x}=0$, red) and non-zonal ($k_{x}\neq0$,
green) parts. The peak of the spectrum at $k\approx1$ is dominated
by zonal modes $k_{x}=0$, but larger scales are also populated by
non-zonal modes, corresponding to large-scale modulations of zonal
jets. In figure~\ref{fig:HMsmallforced}c, the kinetic energy spectrum
is shown at different times (from blue to yellow), where we note the
development of the inverse cascade and the formation of the zonal
peak. In time, the spectrum reaches even larger scales, meaning that
after waiting a sufficient amount of time, a condensate will probably
form \cite{scott:2023}, due to the Rhines scale growing with time
\cite{sukoriansky:2007} in the absence of large-scale friction.
In figure~\ref{fig:HMsmallforced}d, which shows the time evolution
of the zonal velocity profile $\overline{v}_{x}$, the progressive
merging of the jets can be observed. Oscillations are also visible,
related to the modulations of the boundaries of the ZFs.

In short, the zonal pattern that forms in the CHM system is not unidirectional.
Instead, $k_{x}=0$ (zonal) modes appear along with other low $k_{x}$
modes, giving rise to the elliptical/meandering structure of the jets.
Moreover, the width of the jets, which is linked to the Rhines scale
\cite{rhines:75,vallis:1993}, increases with time as they merge,
until they either reach the box size or they equilibrate with large-scale
dissipation \cite{sukoriansky:2007}. Note that the pattern formation
requires a weak forcing (in amplitude), otherwise the system ends up in a turbulent state. If the forcing is not varying sufficiently slowly, 
compared to the linear frequency, the pattern still forms, although less strongly, but not following the wave-turbulence path through resonant interactions (as in \cite{connaughton:15}), and displays more incoherent structures.

\subsection{Instability-driven two-layer QG model}

\begin{figure}[h]
\centering{}\includegraphics[width=1\textwidth]{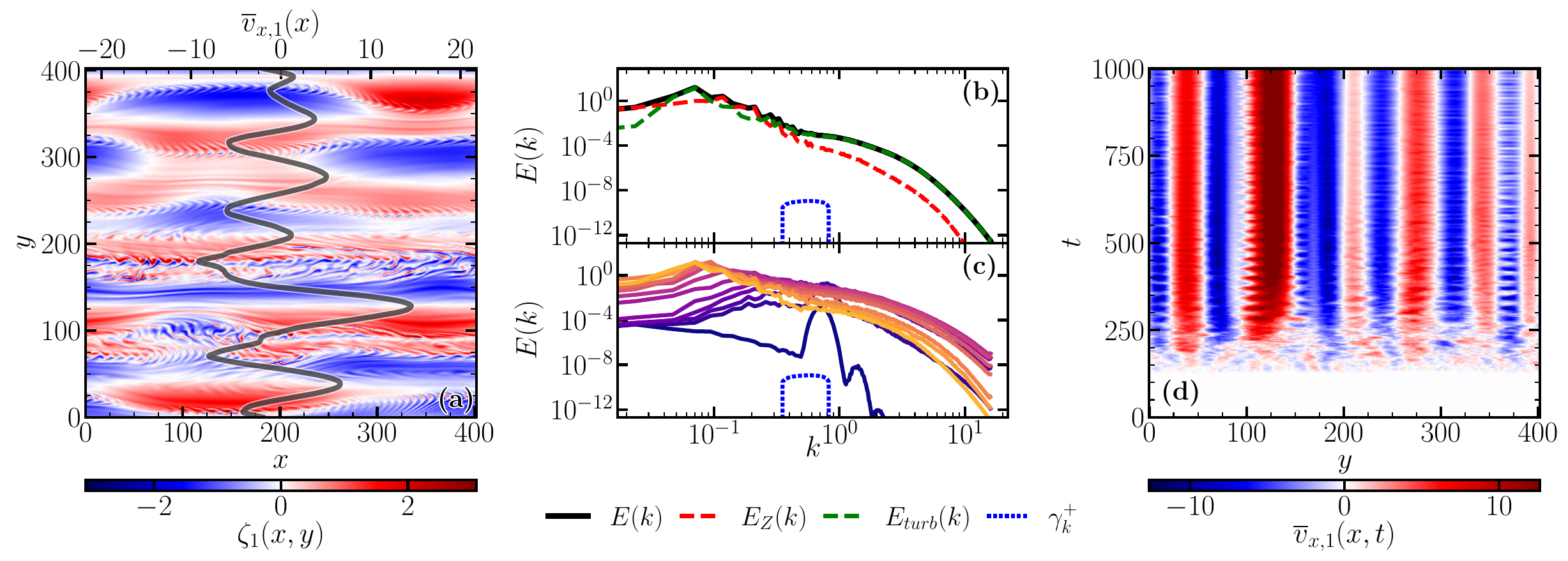}\caption{Two-Layer QG model results for the top layer. (a) Vorticity snapshot
at the final time step and zonal velocity profile (grey line). (b)
Kinetic energy spectrum $E(k)$ (black) time-averaged over $750<t<1000$,
decomposed into zonal $E_{Z}(k)$ (red) and non-zonal $E_{turb}(k)$
(green) parts. The positive part of
the growth rate $\gamma^+_k(k_x)$, which can be found by solving the linear dispersion relation of Eqns.~\eqref{eq:2lqga} and \eqref{eq:2lqgb} (see also \cite{wang:2016}), is in dotted blue; the growth rate is negative otherwise owing to viscous dissipation. (c) Kinetic energy spectrum at different time-steps,
from early stage (blue) to ZFs (yellow). (d) Hovmöller diagram of
the zonal velocity profile.}
\label{fig:QL2G} 
\end{figure}

To investigate layering as the non-linear evolution of an initial
linear instability, we perform a $2048^{2}$ padded resolution DNS
of the two-layer QG model (Eqns.~\eqref{eq:2lqga} and \eqref{eq:2lqgb}).
Using the same normalisation as \cite{wang:2016}, the box size is
$L_{x}\times L_{y}=128\pi\times128\pi$, with $\beta=0.084$, $F_{1}=0.8$,
$F_{2}=0.2$, $U_{1}=1$ and $U_{2}=0$ such that $\Delta U=1$, with
standard minimal diffusion at small scales, with viscosity $\nu=0.05$.
To retain a minimal model, no bottom friction is included. The system
is evolved up to $t=10^{3}$. Results for the top layer (i.e., $i=1$)
are presented in figure~\ref{fig:QL2G}. A time animation of the
system can also be found on the repository page \cite[Movie 2]{guillon:2026b}.

The vorticity snapshot in figure~\ref{fig:QL2G}a shows the system
at the last time step, where we again observe the large-scale elongated
elliptical jet structures, with waves visible at their boundaries.
The kinetic energy spectrum in figure~\ref{fig:QL2G}b shows that
the dominant structures are not purely zonal, with multiple peaks
of the zonal spectrum corresponding to different zonal widths. The
time evolution of the kinetic energy spectrum in figure~\ref{fig:QL2G}c
shows the peak of the linear instability (blue), around $k=0.6$,
which then saturates, resulting in an anisotropic inverse energy transfer
(violet to yellow). From the spatiotemporal evolution of the zonal
velocity profile in figure~\ref{fig:QL2G}d, we observe that the
flows are quite steady and do not undergo any merging after the early
times, with oscillations corresponding to wavelike movement of the
elliptical boundaries of the jets. Notice that the evolution here
is much quicker than that for the CHM system, probably since the instability
makes the system react much faster and stronger than the forcing.

\subsection{Generalised Hasegawa-Mima model}

\begin{figure}[htbp]
\centering{}\includegraphics[width=1\textwidth]{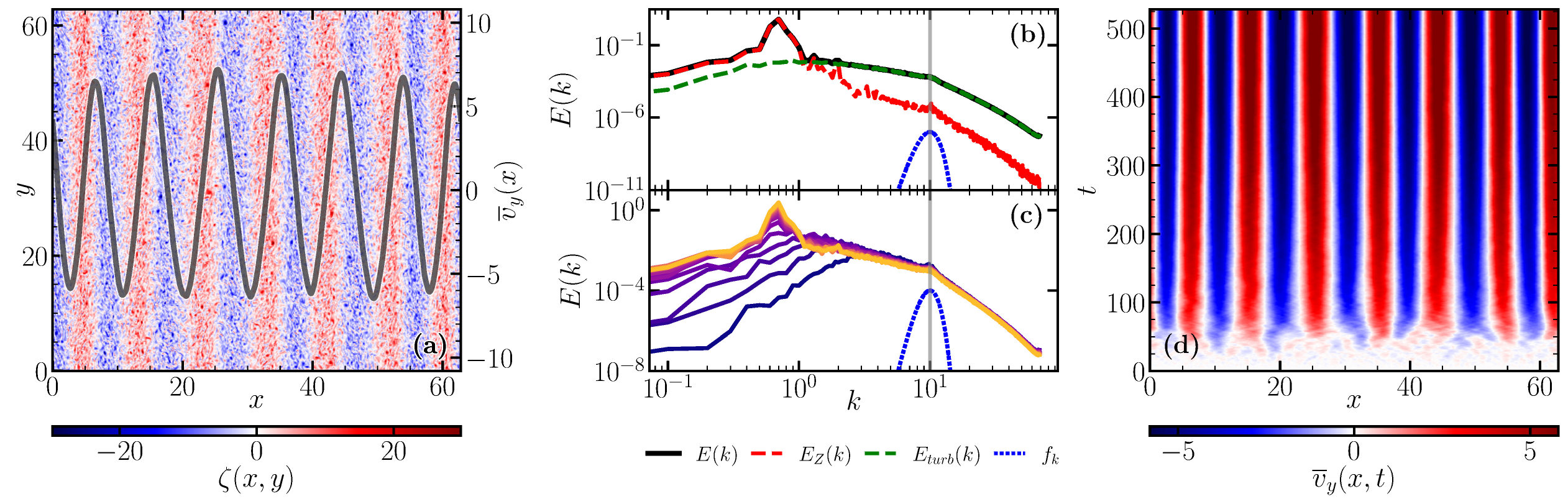}\caption{GHM simulation with small-scale forcing. (a) Vorticity snapshot and
zonal velocity profile (grey line) at the final time step. (b) Kinetic
energy spectrum $E(k)$ (black) time-averaged over $400<t<530$, decomposed
into zonal $E_{Z}(k)$ (red) and non-zonal $E_{turb}(k)$ (green)
parts. The forcing, which is around $k=10$ (vertical grey line),
is in dotted blue. (c) Kinetic energy spectrum at different time-steps,
from early stage (blue) to steady ZFs (yellow). (d) Hovmöller diagram
of the zonal velocity profile.}
\label{fig:GHMsmallforced} 
\end{figure}

We now consider the plasma problem, using the GHM equation, with its
modified zonal response, forced stochastically at small scales. Note
that, as we switch to the plasma convention, the zonal modes correspond
to $k_{y}=0$, or purely $k_{x}$ wave-numbers. We perform a DNS of
the GHM system (Eqn. \ref{eq:ghmend}), with a padded resolution $2048^{2}$,
and a box size of $L_{x}\times L_{y}=20\pi\times20\pi$, with $\kappa=1$,
and $\nu=10^{-3}$, with a forcing centred on $k=10$, of width $\sigma_{k}=1$
and magnitude $A=10^{3}$. The system is evolved up to $t=530$.
The results are shown in figure~\ref{fig:GHMsmallforced}. A time
animation of the system can also be found on the repository page \cite[Movie 3]{guillon:2026b}.

The vorticity snapshot and the zonal velocity profile $\overline{v}_{y}$
at the last time-step (figure~\ref{fig:GHMsmallforced}a) shows an
almost regular pattern, with the jets being completely straight, as
a result of the modified zonal response. This is further illustrated
in figure~\ref{fig:GHMsmallforced}b, where the kinetic energy spectrum
shows that the energy is dominated at large scales by the zonal modes,
which corresponds to the peak at $k_{x}\approx0.8$, with small scales
$k>2$ dominated by 2D turbulence. In figure~\ref{fig:GHMsmallforced}c,
the kinetic energy spectrum is shown at different time-steps (from
blue to yellow), where we see a build-up of the inverse cascade, resulting
in the emergence of ZFs, corresponding to the dominant peak. Here
the ZFs are steady and do not change size during the computation time,
as shown in figure \ref{fig:GHMsmallforced}d.

A key quantity for the formation of ZFs is the so-called Reynolds
power, defined as $P_{Re}\equiv-\int\overline{v}_{y}\partial_{x}R\,\drm x$,
with $R\equiv\langle\widetilde{v}_{x}\widetilde{v}_{y}\rangle_{y}=-\langle\partial_{x}\widetilde{\Phi}\partial_{y}\widetilde{\Phi}\rangle_{y}$
being the Reynolds stress, which represents the power that goes from
2D turbulence to ZFs (i.e., $\drm\mathcal{K}_{Z}/\drm t=P_{R_{e}}$,
where $\mathcal{K}_{Z}=(1/2)\int\overline{v}^{2}_{y}\,dx$ is the
zonal kinetic energy). We observed (not shown here) that for forced
GHM, the Reynolds power stays positive, meaning that ZFs continue
to receive energy. In contrast, the ZFs can actually turn off the
drive in the instability-driven case.

Note that here, the amplitude of the forcing is much higher than for the CHM model, and that the zonal pattern is very robust to increasing the `incoherency' of the forcing (i.e. quickly varying forcing). However, considering a much weaker forcing, the zonal flows are slower to form, they are less strong and tend to merge, resulting in the formation of a condensate. We also tried forcing the GHM system at large scales, using the linear growth rate from the Hasegawa-Wakatani system, which introduces an \textit{ad hoc} linear instability into the system. Zonal flows are generated as usual, but they do not lead to a saturated state in which they dominate, contrary to when the system is forced at small scales. Instead, the flows coexist with strong eddies, especially dipoles, which progressively dominate the system. This is because an \textit{ad hoc} linear instability cannot be turned off by the layering associated with the formation of ZFs.

\section{The Hasegawa-Wakatani system}

\label{sec:HWFG}

In this section, we discuss the formation of zonal flows in the Hasegawa-Wakatani
system, governed by Eqns.~\eqref{eq:hwd} and \eqref{eq:hwv}, driven
by linear instability, where the ratio of the adiabaticity parameter
to the mean density gradient $C/\kappa$ plays the role of the control
parameter. First, we look at the standard case $C/\kappa=1$, which
exhibits stationary ZFs dominating the system. We then study the transition
from 2D turbulence to quasi-1D ZFs as $C/\kappa$ is varied. Finally,
we discuss the saturation mechanism in the system when it is dominated
by ZFs, and how this relates to the layer formation.

\subsection{The standard ($C/\kappa=1$) case.}

\begin{figure}[tbph]
\centering{}\includegraphics[width=1\textwidth]{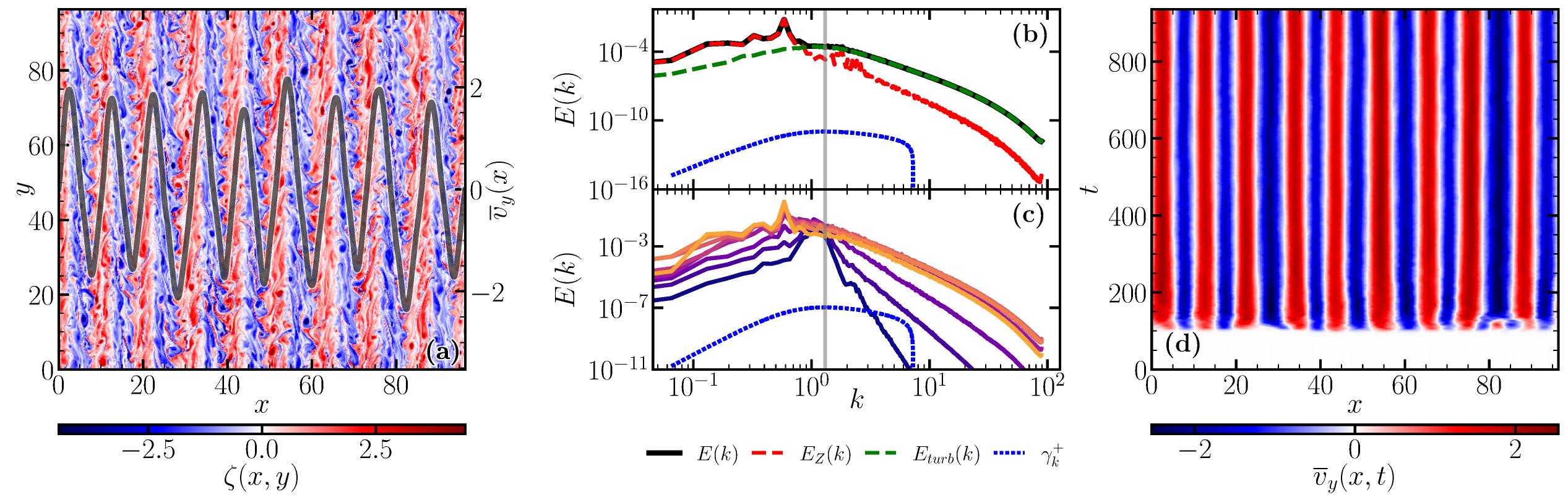}\caption{HW simulation with $C=1$, $\kappa=1$. (a) Vorticity snapshot and
zonal velocity profile (grey line) at the final time step. (b) Kinetic
energy spectrum (black), time-averaged over $475<t<950$, decomposed
into zonal (red) and non-zonal (green) parts. The positive part of
the growth rate, which is maximised at $k_{y0}\approx1.3$ (vertical
grey line) is in dotted blue; the growth rate becomes negative at
small scales owing to viscous dissipation. (c) Kinetic energy spectrum
at different timesteps, from the saturation of the linear instability
(blue) to the steady ZFs (yellow). (d) Hovmöller diagram of the zonal
velocity profile.}
\label{fig:HWC1} 
\end{figure}

We perform a DNS of the Hasegawa-Wakatani system with a padded resolution
$4096^{2}$ and parameters $C=1$ and $\kappa=1$, on a box size $L_{x}\times L_{y}=40\pi/k_{y0}\times40\pi/k_{y0}$,
where $k_{y0}\approx1.3$ is the poloidal wavenumber of the most unstable
mode from the linear dispersion relation~\eqref{eq:freq}, which
acts as a proxy for the energy injection scale. We introduce small-scale
dissipation using standard diffusion terms $\nu\nabla^{2}\widetilde{\zeta}$
and $D\nabla^{2}\widetilde{n}$ respectively for vorticity and density,
with $\nu=D=3.1\cdot10^{-4}$, in order to compensate for energy injection
by the linear instability. Note that no dissipation is applied to
the zonal fields, either at small or large scales, since we observe
that the system is able to converge regardless. The system is evolved
up to $t=950$. The simulation results are presented in figure~\ref{fig:HWC1}.
A time animation of the system can also be found on the repository
page \cite[Movie 4]{guillon:2026b}.

The vorticity snapshot in figure~\ref{fig:HWC1}a shows a system
dominated by quasi-periodic ZFs, similar to the GHM system forced
at small scales, with 2D eddies of sizes similar to the radial width
of the jets. The kinetic energy spectrum (figure~\ref{fig:HWC1}b)
illustrates this final state, where large scales ($k<0.8$) are dominated
by ZFs (dashed red), with the peak at $k\approx0.6$ corresponding
to the mean radial width of ZFs, while smaller scales are dominated
by 2D turbulence, with a peak close to that of the most unstable mode
$k_{y0}$, which is only two times larger than the wavenumber at which the zonal spectrum
peak. The time evolution of the kinetic energy spectrum (figure~\ref{fig:HWC1}c)
shows that the energy, initially concentrated around the most unstable
mode when the linear instability saturates, progressively moves to
the peak of the zonal spectrum.

Finally, the evolution of the zonal profile (figure~\ref{fig:HWC1}d)
shows that the ZFs are extremely steady, with typical width and spacing
of jets that are not dependent on the box size. Comparing the previous
results for CHM to GHM, along with the results on the two-layer QG
model and earlier results on the Hasegawa-Wakatani model with standard
zonal response \cite{pushkarev:13}, we argue that the steadiness
of the jets is a consequence of the modified zonal response above
anything else. Furthermore, in this case, the Reynolds power (not
shown here) is zero in the steady state. Once established, ZFs locally
modify the linear properties of the system. Energy injection by the
instability is weakened and shifted to smaller scales, where it is
removed by enstrophy dissipation.

In a similar vein, it is instructive to note that, in a reduced kinetic
model for trapped ion turbulence in tokamak plasmas, which bears some analogy
with the Rayleigh--Bénard instability in classical fluids, the impact
of the electron response to zonal modes has been studied on a systematic
basis, with a gradual increase from zero (analogous to GHM) to adiabatic
(analogous to CHM) \cite{sarazin:2005, cartier-michaud:2014}. While
the former regime is dominated by steady zonal flows, preventing any
significant cross-field transport, the latter exhibits elongated streamer-like
structures and large transport. These observations give further credit
to the earlier statement about the key role of the
response to zonal modes on the emergence of layering. 

\subsection{The 2D to quasi-1D transition, and the transition scale}

As discussed above, the formation of ZFs in the Hasegawa-Wakatani
system depends on whether or not the control parameter $C/\kappa$
is larger than the critical value $(C/\kappa)_{c}\approx0.1$ \cite{numata:2007,grander:2024,guillon:2025}.
If $C/\kappa<0.1$, the system evolves to one of 2D eddies, with typical
sizes slightly larger than that of the most unstable mode, in a state
that can be characterised as disordered, or `hot', from a thermodynamical
perspective. In contrast, for $C/\kappa>0.1$, one finds the system
dominated by ZFs, as discussed above for $C/\kappa=1$, which becomes
quasi-1D. This regime is more organised and can be seen as a `cold'
crystalline-like structure. For an illustration, we refer, e.g., to
figure 3 from \cite{guillon:2025}, and an animation of the three
regimes can also be found on the repository page \cite[Movie 5]{guillon:2026b}.

\begin{figure}[tbph]
\centering{}\includegraphics[width=\textwidth]{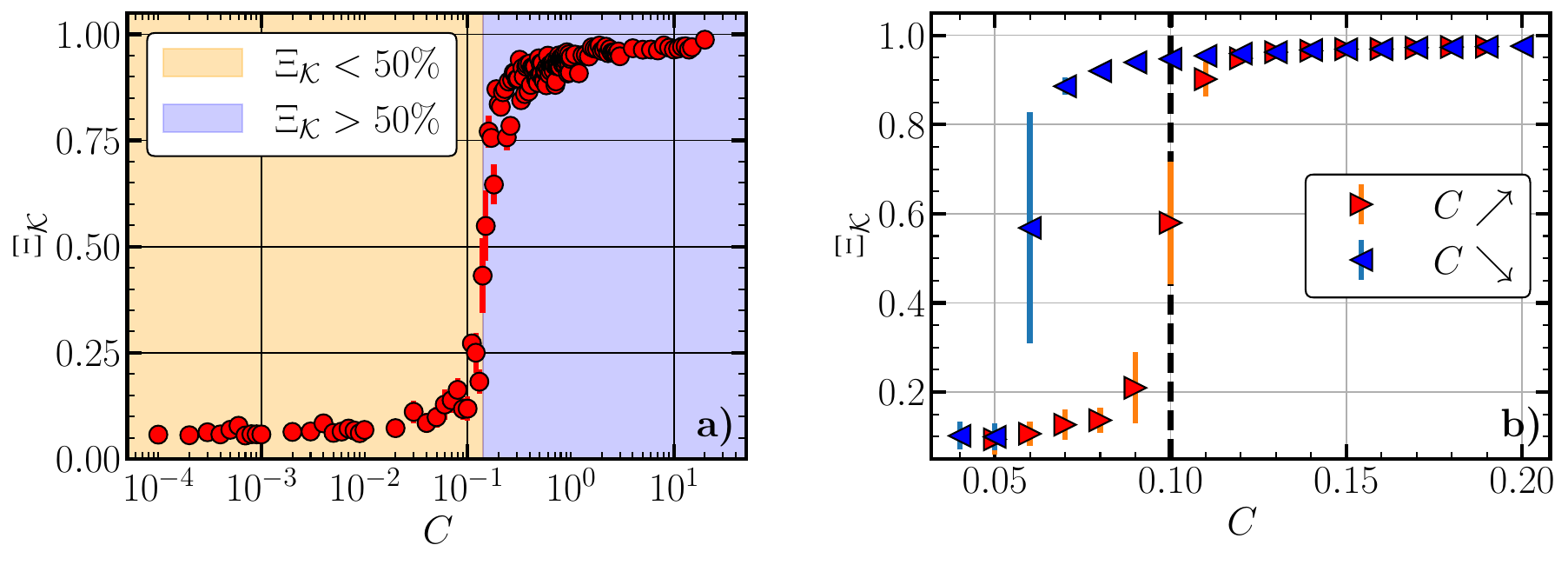}\caption{Transition of the zonal flow level in the HW system on varying $C/\kappa$
(with $\kappa=1$). (a) Individual scans of $C/\kappa$. (b) Hysteresis
loop obtained by slowly increasing (red) and then decreasing (blue)
$C$ in one simulation. The dotted line denotes the transition point
$C/\kappa\approx0.1$.}
\label{fig:HWtranhyst} 
\end{figure}

This bifurcation can hence be viewed as a `phase transition' between
the hot, disordered turbulent regime, and the colder, more organised
ZF dominate state. The transition can then be studied systematically
by quantifying the fraction of zonal to total kinetic energy,
corresponding to the parameter $\Xi_{\mathcal{K}}=\mathcal{K}_{Z}/\mathcal{K}$ (with $0\leq\Xi_{\mathcal{K}}\leq1$), when
scanning the parameter $C/\kappa$. Shown in figure~\ref{fig:HWtranhyst}a,
such a scan exhibits a sharp transition of the order parameter at
approximately $C/\kappa\approx0.1$, which clearly separates the 2D
turbulent state (low zonal flow level, marked in orange) from the zonal-flow-dominated
state (zonal flow level close to $100\%$, marked in blue).

Furthermore, slowly increasing and then decreasing the control parameter
around the transition point $C/\kappa\approx0.1$, while waiting for
a time much larger than the inverse of the instability growth rate, thus performing
an adiabatic transformation, one is able to observe a hysteresis loop
in the transition from the zonal state back to the turbulent state
\cite{guillon:2025}, as shown in figure~\ref{fig:HWtranhyst}b.
Pursuing the analogy with phase transitions, this means that once
zonal flows have been formed, additional latent heat is required in
order to destroy them, somewhat similar, for example, to crystal melting.

While there is still not a complete understanding of the actual mechanism
for the transition and the underlying hysteresis, it can be interpreted
in ways similar to the 3D to quasi-2D transitions \cite{alexakis:2018}
observed in thin layer turbulence \cite{benavides:2017} or in rotating
turbulent flows \cite{seshasayanan:2018,gome:2025}, using here the
existence of a third invariant in $\beta$-plane turbulence, namely
\textit{zonostrophy} \cite{balk:1991,balk:1991b,nazarenko:2009,connaughton:15}.

The transition between 2D turbulence and ZFs can also be discussed
using a spectral perspective on defining the \emph{zonostrophy parameter}
\cite{scott:2012,gurcan:2024} as the ratio $R_{\beta}=k_{\beta}/k_{peak}$
between the energy containing scale $1/k_{peak}$ (i.e., the Rhines
scale in $\beta$-plane turbulence) and the transition scale $1/k_{\beta}$
at which the Rossby/drift wave frequency is equal to the inverse eddy
turnover time. When $R_{\beta}>1$, the energy spectrum peak lies
well in the wave turbulence regime and the system is dominated by
ZFs. On the contrary, for $R_{\beta}<1$ energy is contained in the
2D turbulence scales.

However, in the case of HW, the definition of zonostrophy (or zonostrophy
parameter) is not trivial, since the dispersion relation yields complex
eigenvalues and thus a growth rate (which generally makes it difficult
to apply the weak wave turbulence framework). Furthermore, this argument
does not explain either the sharp transition, or the exact transition
point, the fact that ZFs have a defined size and amplitude (instead
of being a condensate), or the presence of the hysteresis loop. All
these pending questions seem tied up altogether.

\section{Conclusion}

Having studied various forced and instability-driven formulations
of $\beta$-plane turbulence in geophysical and plasma settings, with
different zonal flow responses, we conclude that, while the preference
for layering is universal, the detailed mechanisms of staircase formation,
as well as saturation, are different between forced and instability-driven
systems. But most crucially, the resulting observed pattern seems to depend much more on the nature of the zonal response, rather than on the energy injection mechanism itself.
Nonetheless, there is a common theme in all the models that we have
considered; they all result in PV staircases consisting mainly of
zonal flows.

The general mechanism of layering, when PV is slowly injected, is
that the local PV gradients increase until the system transitions into a high
flux turbulent state, either through a linear instability or a non-linear bifurcation. Then, it relaxes back to the critical state, as
a result of this enhanced flux activity, which is linked to self-organised
criticality/bi-stability through avalanches. While we think of self-organisation
and layering usually in terms of profiles of density or temperature,
in the plasma context, when formulated in terms of PV, vorticity self-organisation
is also key. This allows even a fixed gradient model to `relax' its
PV, through its vorticity evolution, as the only freely evolving channel
for the PV in such a model. In this sense, the differences between
instability-driven and stochastically-forced systems are more qualitative
in that the instability-driven system can turn off the energy injection,
as a result of layering, while the stochastic system needs explicit
sinks in order to saturate properly.

There is also the important effect of the zonal flow response. With
the standard response, the resulting flows appear as very elongated
structures with elliptical jets. In contrast, for the modified zonal
flow response, the resulting flows are really 1D and tend to be very
stationary. This difference can be observed both between the standard
and modified CHM systems, but also between the Hasegawa-Wakatani and the
two-layer QG models. Note that for the CHM model, we obtained these results using a finite Rossby deformation radius $R$. In the limit $R\to +\infty$, the zonal response becomes closer to the modified one, and more straight jets are observed (see, e.g., \cite{scott:2012}).

On considering the saturated state of the Hasegawa-Wakatani system,
with minimal dissipation, we find that the system exhibits a subcritical
zonal bifurcation that resembles a phase transition around $C/\kappa\approx0.1$,
with a hysteresis loop that suggests that, once formed, the zonal
flows require more energy injection to collapse. Note that it could be interesting to investigate whether a similar transition can be found in the two-layer QG model. We further observe
that the generic saturation mechanism, without large-scale friction,
e.g., for a moderate value of $C/\kappa=1.0$, requires both
the suppression of the instability by ZFs and a 2D forward enstrophy
cascade with small-scale dissipation, with a steady state that has
zero net Reynolds power. This happens in a very coordinated way, in
which different parts of the zonal profile either contribute to flattening
of the PV gradient or to generating sufficient shear suppression that
the generated fluctuations are rapidly scattered to small scales,
all happening without generating any momentum flux.

Finally, note that a system driven by a constant gradient of some
quantity (usually the PV itself, but sometimes another background gradient
can play a role), is fundamentally different from a \emph{flux-driven}
system \cite{garbet:1998,gillot:2023,panico:2025,guillon:2025c}. Driven by sources and sinks, such a system can saturate
by relaxing the background gradients that drive small-scale motions,
providing a further mechanism for staircase formation, in addition
to those for forced or instability-driven systems. In that case, close
to the marginal threshold for the layering formation, the feedback
loop between profile relaxation towards the stable state and the feeding
of the source, which increases the mean gradient, can exhibit behaviour
reminiscent of self-organised criticality in sandpile models \cite{bak:1987,hwa:1992,carreras:1996}

\section*{Acknowledgment}
The authors thank the Isaac Newton Institute for Mathematical Sciences, Cambridge, for support and hospitality during the programme `Anti-diffusive dynamics: from sub-cellular to astrophysical scales', which was supported by EPSRC grant EP/R014604/1. This work was granted access to the Jean Zay super-computer of IDRIS
under the allocation AD010514291R2 by GENCI, and has been carried out within the framework of the EUROfusion Consortium, funded by the
European Union \emph{via} the Euratom Research and Training Programme (Grant
Agreement No 101052200 --- EUROfusion) and within the framework of
the French Research Federation for Fusion Studies. Y.S.\ and G.D-P.\ acknowledge that this work was partially supported by a grant from the Simons Foundation. The authors acknowledge stimulating discussions with participants at the 2025 Festival de Th\'eorie in Aix-en-Provence.

\end{document}